\documentclass{epl}
\usepackage{epsfig}
\usepackage{graphicx}
\usepackage{psfig,epsfig}
\usepackage{dcolumn}
\usepackage{bm}

\def\lan{\langle}
\def\ran{\rangle}
\def\e{\mbox{e}}

\title{Glass transition in granular media}
\author{M. Tarzia\inst{1} \and A. de Candia\inst{1} \and 
A. Fierro\inst{1} \and
M. Nicodemi\inst{1} \and
A. Coniglio\inst{1}}
\institute{
  \inst{1} Dipartimento di Fisica, Universit\'{a} di Napoli ``Federico II'',
INFM and INFN, via Cintia, 80126 Napoli, Italy}


\begin{document}

\maketitle

\begin{abstract}
In the framework of schematic hard spheres lattice models for granular media
we investigate the phenomenon of the ``jamming transition''.
In particular, using Edwards' approach, by analytical calculations
at a mean field level,
we derive the system phase diagram and show that ``jamming'' corresponds to
a phase transition from a ``fluid'' to a ``glassy'' phase, observed
when crystallization is avoided.
Interestingly, the nature of such a ``glassy'' phase turns out to be the
same found in mean field models for glass formers.
\end{abstract}

Gently shaken granular media exhibit a strong form of ``jamming'' 
\cite{Knight,Danna,Bideau}, i.e., an exceedingly slow dynamics, 
which shows deep connections \cite{NCH,LN,capri} 
to ``freezing'' phenomena observed 
in many thermal systems such as glass formers \cite{debesti}. 
Although the idea of a unified description of these phenomena 
is emerging \cite{LN}, 
the precise nature of jamming in non-thermal systems and the origin of 
its close connections to glassy phenomena in thermal ones are still open 
and very important issues \cite{OHern}.

Here, we discuss these topics in the framework of the Statistical Mechanics 
of powders introduced by Edwards' \cite{Edwards,e1,fnc} where, to allow 
theoretical calculations, it is assumed that time averages 
of a system subject to some drive (e.g., ``tapping'') coincide with 
suitable ensemble averages over its ``mechanically stable'' states. 
In particular, we consider a schematic model for granular media
recently shown \cite{fnc} to be well described by Edwards' assumption: 
a system of hard spheres under gravity confined on a cubic lattice. 
In this letter we first show that this model subject to a Monte Carlo (MC) 
``tap dynamics'', when crystallization is avoided, has a 
pronounced jamming similar to the one found in 
experiments \cite{Knight,Danna,Bideau}. 
We then discuss the nature of such a form of jamming by analytically solving 
Edwards' partition function of the system at a mean field level, 
by use of a Bethe approximation. 
This approach shows that the present model for granular media 
undergoes a phase transition from a (supercooled) ``fluid'' phase to a 
``glassy'' phase, when its crystallization transition is avoided. 
The nature of such a ``glassy'' phase results to be the same found 
in mean field models for glass formers \cite{Kurchan,Biroli}: a discontinuous 
Replica Symmetry Breaking phase preceded by a dynamical freezing 
point. This finding quantitatively confirms early speculations about the 
structure of jamming in granular media \cite{NCH,Mehta} and clarifies the 
deep connections with glass formers \cite{LN}.


Our schematic model for granular media is a system of monodisperse 
hard-spheres (with radius $a_0=1$) under gravity, constrained to move 
on the sites of a cubic lattice of spacing $a_0$. Its Hamiltonian is:
\begin{equation} 
{\cal H} = {\cal H}_{HC} + mg \sum n_i  z_i
\label{H}
\end{equation}
where $z_i$ is the height of site $i$, $g=1$ is gravity acceleration,
$m=1$ the grains mass, $n_i\in\{0,1\}$ is an occupancy variable 
(absence or presence of a grain on site $i$) 
and ${\cal H}_{HC}(\{n_i\})$ is the hard core term 
preventing the overlapping of nearest neighbors grains \cite{nota_HC}.

$N$ grains are confined in a 3D 
box of linear horizontal size $L$ and height $H$ 
(in our MC simulations $L=12$, $H=20$, $N=288$) between hard walls and 
periodic boundary conditions in the horizontal directions.
They are initially prepared in a random loose stable pack and 
are subject to a dynamics made of a sequence of MC ``taps'' \cite{NCH}:
a single ``tap'' is a period of time,
of length $\tau_0$ (the tap duration), where particles can diffuse
laterally or upwards with a probability $p_{up}\in[0,1/2]$
and downwards with $1-p_{up}$;
when the ``tap'' is off, grains can only move downwards (i.e., $p_{up}=0$)
and reach a blocked state (i.e., one of their ``inherent states'' \cite{fnc}) 
where no one can move downwards without violating the hard core repulsion.
The parameter $p_{up}$ has an effect equivalent to keep the system in contact
(for a time $\tau_0$) with a bath temperature 
$T_{\Gamma}=mga_0/\ln[(1-p_{up})/p_{up}]$ (called the ``tap amplitude'').
Our measurements are performed when the shake is off and the system at rest. 
Time, $t$, is measured as the number of taps applied to the system. 

\begin{figure}[ht]
\vspace{-1.5cm}
\centerline{\hspace{-1.5cm}
\psfig{figure=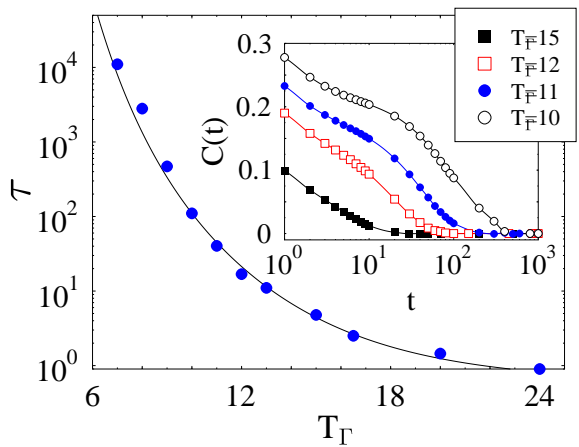,width=9cm,angle=-90}
}
\vspace{-1.8cm}
\caption{{\bf Inset} The normalized site density correlation, $C(t)$, 
is plotted, at stationarity, as a function of the
number of taps, $t$, for the shown ``shaking amplitudes'' $T_\Gamma$ 
(tap duration $\tau_0=100$MC step per site). 
{\bf Main Frame} The characteristic 
relaxation time, $\tau$, is shown as a function of $T_\Gamma$. 
} 
\label{fig_tau}
\end{figure}


The MC tap ``dynamics'' exhibits large relaxation times which grow as the tap 
amplitude decreases. 
In particular, we consider the density correlation function 
$C(t,t_w)=B(t,t_w)/B(t_w,t_w)$, where 
$B(t,t_w)=\sum_i [\lan n_i(t+t_w)n_i(t_w)\ran -\lan n_i(t+t_w)\ran
\lan n_i(t_w)\ran]$. 
At high $T_\Gamma$, for $t_w$ long enough, $C(t,t_w)$ has a 
time translation invariant behavior, i.e., $C(t,t_w)=C(t)$.
We do not consider here low values of $T_\Gamma$ as the model tends to
crystallization \cite{nota_cr}. 
The relaxation time, $\tau(T_\Gamma)$, plotted in Fig.\ref{fig_tau}, has been
obtained by fitting $C(t)$ as an exponential function at long time $t$.
Even though we are at comparatively 
high $T_\Gamma$ values, an Arrhenius law (shown in the picture) or 
a Vogel-Tamman-Fulcher law fits the data. These  properties 
are very similar to those 
found in more refined models \cite{fnc,NCH,cdfnt} 
and correspond to recent experimental results on granular media 
\cite{Danna,Bideau}, which are found to exhibit a glassy behavior  
\cite{debesti}. 

\begin{figure}[ht]
\centerline{
\psfig{figure=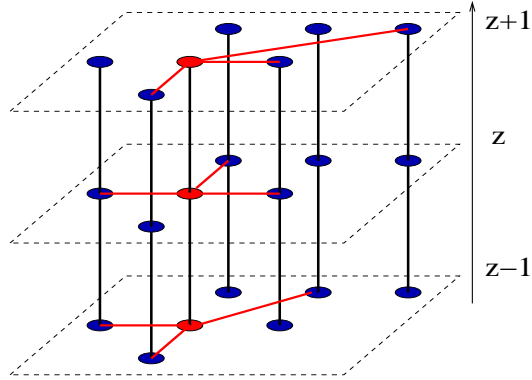,width=7cm,height=5cm,angle=0}
}
\vspace{-.3cm}
\caption{In our mean field approximation, the hard spheres describing 
the system grains are located on a Bethe lattice, sketched in the figure, 
where each horizontal layer is a random graph of given connectivity. 
Homologous sites on neighboring layers are also linked and the overall 
connectivity, $c$, of the vertices is $c\equiv k+1=5$.} 
\label{Blattice}
\end{figure}

The nature of the glassy region is, anyway, very difficult to be numerically 
established and further insight can be obtained by analytical treatments. 
This can be accomplished in Edwards' approach \cite{Edwards,e1,fnc} 
which we now assume to hold in our model, at least as a good approximation. 
This assumption is grounded on some recent results \cite{fnc} showing 
that ``time averages'' of the system observables over the ``tap'' dynamics 
coincide with those over an ensemble average where only 
``mechanically stable'' states, (i.e., those where the system is found 
at rest), are counted. More specifically, the weight of a given state 
$r$ is \cite{fnc}: $\e^{-\beta {\cal H}(r)}\cdot \Pi_r$,  
where $T_{conf} = \beta^{-1}$ is a thermodynamic parameter, 
called ``configurational temperature'', characterizing the distribution
($T_{conf}$ can be related to the shaking amplitude $T_\Gamma$, 
for instance, from the equality between the time average of the energy, 
${\overline e} (T_\Gamma)$, 
and the ensemble average, $\lan e\ran(T_{conf})$). 
The operator $\Pi_r$ selects mechanically stable states: 
$\Pi_r=1$ if $r$ is ``stable'', else $\Pi_r=0$. 
The system partition function \'a la Edwards is thus the following \cite{fnc}: 
\begin{equation}
Z=\sum_{r} \e^{-\beta {\cal H}(r)}\cdot \Pi_r
\label{Z}
\end{equation}
where the sum runs over all microstates $r$. 

Since the exact calculation of $Z$ for the above lattice model 
is hardly feasible, we now discuss a mean field theory 
(see \cite{Biroli,MP} and ref.s therein) based on a random graph 
version of such a lattice which is sketched in Fig.\ref{Blattice}.
More specifically we consider a 3D lattice box with $H$ horizontal layers 
(i.e., $z\in\{1,...,H\}$) occupied by hard spheres. 
Each layer is a random graph of given connectivity, $k-1$ 
(we take $k=4$). Each site in layer $z$ is also 
connected to its homologous site in $z-1$ and $z+1$ 
(the total connectivity is thus $k+1$). 
The Hamiltonian is the one of eq.(\ref{H}) plus a chemical potential 
term to control the overall density. Hard Core repulsion prevents 
two connected sites to be occupied at the same time. 
In the present lattice model we adopt a simple definition of ``mechanical 
stability'': a grain is ``stable'' if it has a grain underneath. 
For a given grains configuration 
$r=\{n_i\}$, the operator $\Pi_r$ has thus a simple expression: 
$\Pi_r =\lim_{K\rightarrow\infty}\exp\left\{-K
{\cal H}_{Edw}\right\}$ 
where ${\cal H}_{Edw}=
\sum_i \delta_{n_i(z),1}\delta_{n_i(z-1),0}\delta_{n_i(z-2),0}$ 
(for clarity, we have shown the $z$ dependence in $n_i(z)$). 


The local tree-like structure of our lattice allows to write down 
iterative equations \'a la Bethe for the probability of fields acting on the 
lattice sites \cite{nota_loops}. 
In the ``cavity method'' \cite{MP}, the recurrence equations are found 
by iteration of the lattice structure where $k$ ``branches'' 
(i.e., graphs where a root site, denoted by $j\in\{1,...,k\}$, has only 
$k$ neighbors) are merged to a new site $i$, leading to a lattice with 
the same structure as before but with one more site. 
Three kinds of ``branches'' exist here:
``up'' (resp. ``down'') branches where the root site has $k-1$ neighbors 
on its same layer and one in the upper (resp. lower) layer; 
and ``side'' branches where the root has $k-2$ neighbors on its layer, 
one in the upper and one in lower layer. 
Correspondingly, three kinds of merging are possible where: 
an ``up'' (resp. ``down'') branch with root at height $z+1$ (resp. $z-1$) 
and $k-1$ ``side'' branches with root at height $z$ merge into a new 
``up'' (resp. ``down'') branch with root in site $i$ at height $z$;
an ``up'', a ``down'' and $k-2$ ``side'' branches merge into a new 
``side'' branch. 

\begin{figure}[ht]
\vspace{-1.4cm}
\centerline{\hspace{-2cm}
\psfig{figure=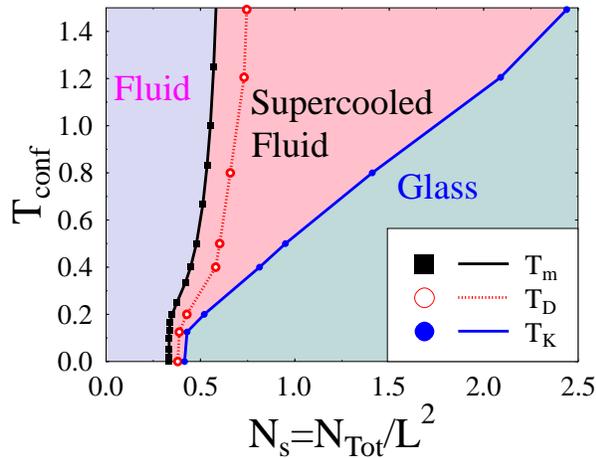,width=9cm,angle=-90}
}
\vspace{-1.5cm}
\caption{The system mean field phase diagram is plotted in the plane 
of its two control parameters $(T_{conf},N_s)$: $T_{conf}$ is Edwards' 
``configurational temperature'' and $N_s$ the average 
number of grains per unit surface in the box. 
At low $N_s$ or high $T_{conf}$, the system is found in a fluid phase. 
The fluid forms a crystal below a melting transition line $T_m(N_s)$. 
When crystallization is avoided, the ``supercooled'' (i.e., metastable) 
fluid has a thermodynamic phase transition, at a point $T_K(N_s)$, 
to a Replica Symmetry Breaking ``glassy'' phase with the same structure 
found in mean field theory of glass formers. In between $T_m(N_s)$ 
and $T_K(N_s)$ a dynamical freezing point, $T_D(N_s)$, is located, where 
the system characteristic time scales diverge.}
\label{MFPD}
\end{figure}

The partition function of the new branch ending in site $i$ can be 
recursively written in terms of the partition functions of the merged 
branches. 
Define $Z_{0,s}^{(i,z)}$ and $Z_{1,s}^{(i,z)}$ the partition functions 
of the ``side'' branch restricted respectively to configurations 
in which the site $i$ is empty or filled by a particle; analogously, 
$Z_{1,u}^{(i,z)}$ and $Z_{0,u}^{(i,z)}$ (resp. $\overline{Z}_{0,u}^{(i,z)}$) 
are the partition functions of the ``up'' branch restricted to 
configurations in which the site $i$ is filled or empty with the upper site 
filled (resp. empty); finally $Z_{1,d}^{(i,z)}$ and $Z_{0,d}^{(i,z)}$ 
(resp. $\overline{Z}_{0,d}^{(i,z)}$) are those of the ``down'' branch 
when site $i$ is filled or empty with the upper site empty (resp. filled). 
For more details see \cite{cdfnt}. 
It is convenient to introduce five local ``cavity fields'' 
$h_s^{(i,z)}$, $h_u^{(i,z)}$, $g_u^{(i,z)}$, $h_d^{(i,z)}$ and $g_d^{(i,z)}$ 
defined by: 
$e^{\beta h_s^{(i,z)}} = Z_{1,s}^{(i,z)}/Z_{0,s}^{(i,z)}$; 
$e^{\beta h_u^{(i,z)}} = Z_{1,u}^{(i,z)}/\overline{Z}_{0,u}^{(i,z)}$; 
$e^{\beta g_u^{(i,z)}} = Z_{0,u}^{(i,z)}/\overline{Z}_{0,u}^{(i,z)}$; 
$e^{\beta h_d^{(i,z)}} = Z_{1,d}^{(i,z)}/\overline{Z}_{0,d}^{(i,z)}$; 
$e^{\beta g_d^{(i,z)}} = Z_{0,d}^{(i,z)}/\overline{Z}_{0,u}^{(i,z)}$. 
In these new variables the recursion relations 
are more easily written \cite{nota_bc}:
\begin{eqnarray} \label{ricorrenza}
\nonumber
e^{\beta h_s^{(i,z)}} & = & e^{\beta(\mu-mgz)}\left[\prod_{j=1}^{k-2}
(1+e^{\beta h_s^{(j,z)}})^{-1}\right]
(1+e^{\beta g_u^{(i,z+1)}})\times \\
\nonumber
& &
[1+e^{\beta h_d^{(i,z-1)}}+e^{\beta g_d^{(i,z-1)}}+e^{\beta h_d^{(i,z-1)}
+\beta h_u^{(i,z+1)}}]^{-1}\\
\nonumber
e^{\beta h_u^{(i,z)}} & = & e^{\beta(\mu-mgz)}
( 1+e^{\beta g_u^{(i,z+1)}} ) \prod_{j=1}^{k-1}
(1+e^{\beta h_s^{(j,z)}})^{-1}
\\
e^{\beta g_u^{(i,z)}} & = & e^{\beta h_u^{(i,z+1)}} \\
\nonumber
e^{\beta h_d^{(i,z)}} & = & e^{\beta(\mu-mgz)} 
e^{-\beta h_d^{(i,z-1)}} \prod_{j=1}^{k-1}
(1+e^{\beta h_s^{(j,z)}})^{-1} \\
\nonumber
e^{\beta g_d^{(i,z)}} & = & (1+e^{\beta g_d^{(i,z-1)}}) 
e^{-\beta h_d^{(i,z-1)}} ~ .
\end{eqnarray}
From the iterative solution of these equations it is possible to compute 
the system free energy 
\cite{MP,cdfnt}. Fig.\ref{MFPD} shows the system phase diagram in the plane 
of the two control parameters, $(T_{conf},N_s)$, where $N_s$ is the 
number of grains per unit surface in the box.

\begin{figure}[ht]
\vspace{-1.4cm}
\centerline{\hspace{-2cm}
\psfig{figure=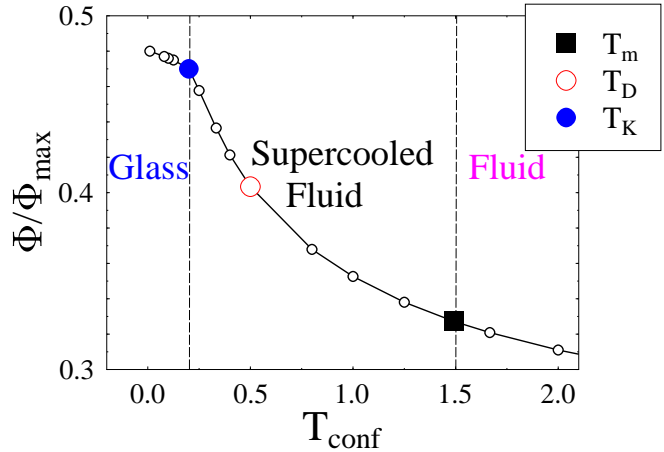,width=9cm,angle=-90}
}
\vspace{-1.5cm}
\caption{For a system with a given number of grains (i.e., a given $N_s$), 
the overall number density, $\Phi\equiv N_s/2\lan z\ran$ ($\lan z\ran$ is the 
average height), calculated in mean field approximation is plotted 
as a function of $T_{conf}$; $\Phi(T_{conf})$ has a shape 
very similar to the one observed in the ``reversible regime'' of 
tap experiments and MC simulations of the cubic lattice model for 
$\Phi(T_{\Gamma})$. 
The location of the glass transition, $T_K$ (filled circle), corresponds 
to a cusp in the function $\Phi(T_{conf})$. 
The passage from the fluid to supercooled fluid is $T_m$ (filled square).
The dynamical crossover point $T_D$ is found around 
the flex of $\Phi(T_{conf})$ and well corresponds to the position 
of a characteristic shaking amplitude $\Gamma^*$ found in experiments 
and simulations 
where the ``irreversible'' and ``reversible'' regimes approximately meet.}
\label{phi_T}
\end{figure}

At low $N_s$ or high $T_{conf}$ a fluid-like phase is found, 
characterized by a homogeneous Replica Symmetric (RS) solution 
(in Replica Theory terminology) of the recursion equations (\ref{ricorrenza}), 
in which only 
one pure state exists and the local fields are the same for all the sites 
of the lattice (translational invariance).
For a given $N_s$, by lowering $T_{conf}$ (see Fig.s \ref{MFPD},\ref{phi_T}), 
a phase transition to a crystal phase (an RS solution 
with no space translation invariance) is found at $T_m$ \cite{nota_2ndo}. 
Notice that the fluid phase still exists below $T_m$ as a metastable 
phase corresponding to a supercooled fluid found when crystallization 
is avoided. 

The above RS fluid solution, however, is not appropriate to describe 
the high $N_s$ or low $T_{conf}$ region 
which is dominated by the presence of a large number of local minima of 
the free energy where the fields may fluctuate \cite{MP}.
This situation is characterized by non-trivial probability distributions 
for the local fields on each layer $z$: 
${P}_{z}^{u} (h_u, g_u)$, ${P}_{z}^{s} (h_s)$ 
and ${P}_{z}^d (h_d, g_d)$. 
Within the one-step replica symmetry breaking (1RSB) ansatz of the 
cavity method \cite{MP,nota_ti}, the recursion relations for the fields 
are replaced by self consistent integral equations for the distribution 
of the local fields:
\begin{eqnarray}
P_{z}^{u} (h_u^z, g_u^z) &=& C_1
\int \prod_{j=1}^{k-1}
\left[ d h_{s}^{(j,z)} P_{z}^{s} (h_s^{(j,z)}) \right] ~ \times 
\left [d h_{u}^{(i,z+1)} d g_{u}^{(i,z+1)} P_{z+1}^{u} (h_u^{(i,z+1)}, g_u^{(i,z+1)})
\right]   
\nonumber \\
&& 
\times \; \; \delta (h_{u}^{z} - h_u^{(i,z)}) \delta (g_{u}^{z} - g_u^{(i,z)})
\exp (-\beta m \Delta F_u^z),
\end{eqnarray}
\noindent
where ${C}_1$ 
is a normalization constant, $h_u^{(i,z)}$, $g_u^{(i,z)}$
are the local fields defined by the eqs.(\ref{ricorrenza}), $\Delta F_u^z$
is the free energy shifts in the merging process \cite{MP,cdfnt} 
and $m$ is the usual 1RSB parameter to be obtained by maximization
of the free energy with respect to it. Analogous equations are found for 
$P_{z}^{s} (h_s)$ and $P_{z}^d (h_d, g_d)$.
We have solved all these equations iteratively, by discretizing the 
probability distributions, until the whole procedure converged.

A non trivial solution of the 1RSB equations appears 
for the first time at a given temperature $T_D(N_s)$, signaling the 
existence of an exponentially high number of pure states. 
In mean field theory $T_D$ is interpreted 
as the location of a purely dynamical transition as in 
Mode Coupling Theory, but in real 
systems it might correspond just to a crossover in the dynamics 
(see \cite{Kurchan,Biroli,Toninelli} and ref.s therein). 
The 1RSB solution becomes stable at a lower point $T_K$, where 
a thermodynamic transition from the supercooled fluid 
to a 1RSB glassy phase takes place (see Fig.\ref{MFPD}) in a scenario 
\'a la Kauzmann with a vanishing complexity of pure states 
(which stays finite for $T_K<T<T_D$). 

The results of these calculations, summarized in the phase diagram of 
Fig.\ref{MFPD}, are further illustrated in Fig.\ref{phi_T}: 
in a system with a given number of grains (i.e., a given $N_s$), the 
overall number density, $\Phi$, is plotted as a function of $T_{conf}$ (here 
by definition $\Phi\equiv N_s/2\lan z\ran$, where $\lan z\ran$ is the 
average height). The shown curve, $\Phi(T_{conf})$, is the equilibrium 
function here calculated. It has a shape very similar to the one 
observed in tap experiments \cite{Knight,Bideau}, 
or in MC simulations on the cubic lattice (see also \cite{NCH}), 
where the density is plotted as a 
function of the shaking amplitude $\Gamma$ (along the 
so called ``reversible branch''). 


Summarizing, in the present mean field scenario of a granular medium 
with $N_s$ particles per surface, in general, 
at high $T_{conf}$ (i.e. high shaking amplitudes) 
a fluid phase is located (see Fig.\ref{MFPD}). 
By lowering $T_{conf}$, a phase transition to a crystal phase is found 
at $T_m$. However, when crystallization is avoided, the fluid phase 
still exists below $T_m$ as a metastable phase corresponding to a 
supercooled fluid. 
At a lower point, $T_D$, an exponentially high number of new 
metastable states appears, interpreted, at a mean field level, 
as the location of a purely dynamical transition, which in real system 
is thought to correspond just to a dynamical crossover. 
Finally, at a even lower point, $T_K$, the supercooled fluid has a 
genuinely thermodynamics discontinuous phase transition to 
glassy state. MC simulations of a cubic lattice model show indeed the 
divergence of the relaxation time $\tau$ at low ``shaking'' amplitudes 
(see Fig.\ref{fig_tau}). 
The structure of the glass transition of the present model for granular media, 
obtained in the framework of Edwards' theory 
is the same found in the glass transition of the $p$-spin glass and 
in other mean field models for glass formers \cite{Kurchan,Biroli}.

Work supported by MIUR-PRIN 2002, MIUR-FIRB 2002, CRdC-AMRA, INFM-PCI.


\begin{thebibliography}{100}

\bibitem{Knight} J.B. Knight {\em et al.}, 
Phys. Rev. E {\bf 51}, 3957 (1995). E.R. Nowak {\em et al.}, 
Phys. Rev. E {\bf 57}, 1971 (1998).

\bibitem{Danna} G. D'Anna and G. Gremaud, Nature {\bf 413}, 407 (2001).

\bibitem{Bideau} P. Philippe and D. Bideau, 
Europhys. Lett. {\bf 60}, 677 (2002).

\bibitem{NCH} M. Nicodemi, A. Coniglio, H.J. Herrmann,
Phys. Rev. E {\bf 55}, 3962 (1997). M. Nicodemi and A. Coniglio,
Phys. Rev. Lett. {\bf 82}, 916 (1999). 

\bibitem{LN} A.J. Liu and S.R. Nagel, Nature {\bf 396}, 21 (1998).

\bibitem{capri} {\em ``Unifying concepts in granular media and glasses''}, 
(Elsevier Amsterdam, in press), 
Edt.s A. Coniglio, A. Fierro, H.J. Herrmann, M. Nicodemi.

\bibitem{debesti} M.D. Ediger, C.A. Angell, S.R. Nagel, J. Phys. Chem. 
{\bf 100}, 13200 (1996). P.G. Debenedetti and  F.H. Stillinger, Nature 
{\bf 410}, 259 (2001).

\bibitem{OHern} C.S. O'Hern, S.A. Langer, A.J. Liu, S.R. Nagel, 
Phys. Rev. Lett. {\bf 86}, 111 (2001). 
C.S. O'Hern, L.E. Silbert, A.J. Liu, S.R. Nagel, {\em cond-mat/0304421}. 

\bibitem{Edwards}  S.F. Edwards and R.B.S. Oakeshott, 
Physica A {\bf 157}, 1080 (1989).
A. Mehta and S.F. Edwards, 
Physica A {\bf 157}, 1091 (1989).
S.F. Edwards, in {\em Current Trends in the physics of
Materials}, (Italian Phys. Soc., North Holland, Amsterdam, 1990).

\bibitem{e1}  M. Nicodemi, Phys. Rev. Lett. {\bf 82}, 3734 (1999). 
A. Barrat {\em et al.}, Phys. Rev. Lett. {\bf 85}, 5034 (2000). 
J.J. Brey, A. Prados, B. S\'{a}nchez-Rey, Physica A {\bf 275}, 310 (2000).
D. S. Dean and A. Lef\`{e}vre, Phys. Rev. Lett. {\bf 86}, 5639 (2001). 
H. A. Makse and J. Kurchan, Nature {\bf 415}, 614 (2002).
J. Berg, S. Franz and M. Sellitto, Eur. Phys. J. B {\bf 26}, 349 (2002). 
G. De Smedt, C. Godreche, J.M. Luck, Eur. Phys. J. B {\bf 32}, 215-225 (2003).
G. Tarjus, P. Viot, {\em cond-mat/0307267}. 

\bibitem{fnc} A. Coniglio and M. Nicodemi, Physica A {\bf 296}, 451 (2001).
A. Fierro, M. Nicodemi and A. Coniglio, Europhys. Lett. {\bf 59}, 642 (2002);
Europhys. Lett. {\bf 60}, 684 (2002); Phys. Rev. E {\bf 66}, 061301 (2002).

\bibitem{Kurchan} L. Cugliandolo and J. Kurchan, 
Phys. Rev. Lett. {\bf 71}, 173 (1993). J. Kurchan, {\em cond-mat/9812347}; and
in ``Jamming and Rheology'',
A.J. Liu and S.R. Nagel Eds.,
Taylor and Francis,  London (2001).

\bibitem{Biroli} G. Biroli and M. M\'ezard, 
Phys. Rev. Lett. {\bf 88}, 025501 (2002). 

\bibitem{Mehta} A. Mehta and J. Berg, Europhys. Lett. {\bf 56}, 784 (2001).

\bibitem{nota_HC} ${\cal H}_{HC}(\{n_i\})=J\sum_{\lan ij\ran }n_in_j$, with
$J\rightarrow\infty$.

\bibitem{nota_cr} Crystallization can be avoided, without altering the 
overall scenario, by introduction of some degree of polidispersity \cite{fnc}, 
or by considering less coarse grained lattice models \cite{fnc,NCH,cdfnt}. 

\bibitem{cdfnt} A. Coniglio, A. de Candia, A. Fierro, M. Nicodemi, 
M. Tarzia, in preparation. 

\bibitem{MP} M. M\'ezard and G. Parisi, Eur. Phys. J. B {\bf 20}, 217 (2001). 

\bibitem{nota_loops} Locally the graph has a tree-like structure but there 
are loops of order $\ln N$ insuring geometric frustration. 

\bibitem{nota_2ndo} In the present model the ``melting'' transition is 
continuous. This pathology can be cured as explained in \cite{cdfnt} or 
in M. Weigt and A.K. Hartmann, Europhys. Lett. {\bf 62}, 533 (2002).

\bibitem{nota_bc} The boundary conditions are insured by adding 
two auxiliary planes at height $z=-1$ and $z=H+1$ 
where all sites are empty but stable (in the Edwards' sense).

\bibitem{nota_ti} Since the glassy phase is expected to be translational 
invariant, we work in the factorized case in which the probability 
distributions at a given height are equal for all the sites of the layer.

\bibitem{Toninelli} C. Toninelli, G. Biroli, D.S. Fisher, 
{\em cond-mat/0306746}.


\end{thebibliography}
\end{document}